\documentclass[12pt]{article}
\begin{document}
\title{THE COUNTERINTUITIVE UNIVERSE}
\author{B.G. Sidharth\\
International Institute for Applicable Mathematics \& Information Sciences\\
Hyderabad (India) \& Udine (Italy)\\
B.M. Birla Science Centre, Adarsh Nagar, Hyderabad - 500 063 (India)}
\date{}
\maketitle
\begin{abstract}
The major advances in physics have been through counterintuitive breakthroughs-- ideas that seemed to go against prevailing convictions. In the twentieth century the Special and General Theory of Relativity and Quantum Mechanics have provided very good examples of this process. However, twentieth century physics has led to an impasse, one of the most important unsolved problems being the unification of gravitation with other interactions. This has led to another sacred tenet of physics, viz., spacetime as a smooth manifold, being questioned in latest approaches, be it Quantum Super Strings or other Quantum Gravity and similar theories. In particular spacetime described by an underpinning of Planck scale oscillators is investigated.
\end{abstract}
\section{Introduction}
The greatest breakthroughs in our concepts of physics, and science in general, have been counterintuitive. Broadly speaking, intuitive concepts are those which are in conformity with our ideas of the universe based on experience gathered till that time. It is difficult to break out of this mould-- generally such a departure is forced upon us due to the inadequacy of the state of our knowledge to describe new phenomena.\\
One of the greatest conceptional breakthroughs of this kind was that of the ancient Indian philosopher, Kanada. Sometime around the seventh century B.C. he postulated that all material objects were made up of ultimate atomic constituents in constant vibration. A few centuries later, atomism was also postulated in the Greek world, probably independently. But these atoms unlike those of Kanada, were not in constant vibration. Though brilliant, these concepts were much too counterintuitive to be accepted wholeheartedly. They remained in the fringes of speculation and philosophy.\\
Another dramatic counterintuitive breakthrough, which however survived, was the observation that the earth which looked flat everywhere, was actually round. Equally radical was the suggestion that the solid earth was infact not at the centre of the universe, but rather was spinning on its axis and whirling round the sun. Though such contra ideas were put forward more than two thousand years ago, they could not be accepted by the people at large or even scholars. Numerous arguments were put forth over the centuries to prove ``decisively'' that the earth could not be travelling or rotating.\\
We had to wait for nearly two thousand years before the next breakthrough. This had to do with the concept of the universe-- the earth was at the centre surrounded by material transparent spheres each of which rotated, carrying objects like the sun, the moon, the planets and the stars. Such material spheres were necessary because, we would have otherwise had to explain why the moon, or the sun for that matter, doesn't crash down. Such a question had been asked for thousands of years.\\
It is interesting to note that Newton's answer to this puzzle was, in a sense the opposite to what one might naively expect. (As we will see, that came much later from Einstein!). There was no force counterbalancing the tendency to crash downwards. There was only one force, the force of gravitation which in the case of the earth-moon system, actually pulled the moon towards the earth. Yet the moon did not crash because of the mechanics-- because of its motion round the earth. The spin and orbit of the earth were also settled during this period in which the old Greek Astronomy was finally thrown overboard.\\
Newton's laws of gravitation and mechanics held sway for just about a tenth of the length of the reign of the Greek model. Newtonian space was distinct from time. It provided an absolute  background or platform on which the drama of the universe, the motions of bodies and projectiles took place. Matter, forces, energy and the like were actors acting out in time. So also the law of gravitation was an action at a distance theory. Every material particle exerted the force of gravitation instantly on every other material particle.\\
In the next century, Coulomb discovered the law of electric, more precisely electrostatic interaction. It had the same form of an inverse square dependence on distance, as gravitation. This too was an action at a distance force. While the action at a distance gravitational law worked satisfactorily, in the nineteenth century the Coulomb law encountered difficulties when it was discovered by Ampere, Faraday and others that moving charges behaved differently. The stage was being set for Maxwell's electrodynamics.\\
Maxwell could unify the experimental laws of Faraday, Ampere and others in a Field Theory \cite{jackson}. Already in the seventeenth century itself Olaf Romer had noticed that light travels with a finite speed and does not reach us instantly. He could conclude this by observing the eclipses of the satellites of Jupiter. Christian Huygens took the cue and described the motion of light in the form of waves. The analogy with ripplies moving outwards on the surface of a pool was clear.\\
Maxwell utilised these ideas in interpreting the experimentally observed laws of electricity and magnetism. Thus a moving charge would cause a ripple in an imaginary medium or field, and that ripple would propagate further till it hit and acted upon another charge. This was a dramatic departure from the action at a distance concept because the effect of the movement of the charge would be felt at a later time by another charge. Maxwell even noticed that the speed at which these disturbances would propagate through the field was the same as that of light. Already the stage was being set for Einstein's Theory of Special Relativity \cite{einstein}.\\ 
Primarily this had to do with the speed of light which seemed to be the same in all frames. How could that be? This seemed to contradict our everyday experience. Einstein adopted this counterintuitive principle as a postulate and the new concept of spacetime emerged. There were other such dramatic new results such as the relativity of simultaneity or spacetime intervals.\\
So obscure were these ideas that Einstein was eventually awarded the Nobel Prize for his work in topics like the Photoelectric Effect, rather than the Special Theory of Relativity. Even so, it must be mentioned that there is an earlier formulation of the old action at a distance theory due to Fokker and others which resembled closely Maxwell's Field Theory, in mathematical form.\\
At this stage it was clear that two closely related concepts were important and entered the realm of our intuitive thought-- locality and causality. We will return to this shortly but broadly what is meant is that parts of the universe could be studied in isolation and further, that an event at a point $A$ cannot influence an event at a point $B$ which cannot be reached by a ray of light during this interval. Roughly speaking, all events within this light radius would be causally connected, but not so events beyond this radius.\\
Till then Newton's gravitation held its ground. There was an ``actual force'' of attraction between any two material objects. Einstein then conceived of his elevator experiment. If one were in an elevator whose cable had snapped for example, he would be in free fall and would not experience gravity till he hit the ground. Conversely if he were in an elevator floating deep in outer space without any trace of gravity, and the elevator was then given an acceleration equal to that of the falling body on the earth, this observer would then feel in the opposite direction the forceof gravity, exactly as on the earth.\\
From this point of view, there is a counterintuitive conclusion-- gravity could be destroyed or created through suitable accelerations. It wasn't a ``real'' force. This lead to Einstein's General Theory of Relativity in which gravitation is an effect of the curvature of spacetime.\\
There was one other concept that Einstein introduced in the belief that the constituents of the universe were static. It was essentially the old problem of Newton and the apple. He postulated a repulsive force via a cosmological constant-- the twentieth century counterpart of the Greek crystalline spheres. Then around 1920 it was discovered that the universe is expanding, and Einstein retracted, dubbing it the greatest blunder of his life. As we will briefly see, we may have to resurrect this force again in the twenty first century.\\
In Einstein's theories, spacetime was no longer a passive background, but it actually participated in the processes. Nevertheless all this could be called Classical Physics.\\
The advent of Quantum Theory was another counterintuitive leap away from old ideas. In the solar system, planets could orbit the sun with any arbitrary energy in principle. Within the atom however electrons could orbit only with certain handpicked energies and nothing else. An electron could however jump from one energy level to another, absorbing or emitting a discrete amount or Quantum of energy. Thus energies were no longer arbitrary. While all this was constructed to explain otherwise inexplicable observations, nevertheless Quantum Theory succeeded in fudging its ad hoc character and providing a more fundamental justification for itself over the years.\\
A crowning achievement of Quantum Theory was the Standard Model of Weinberg, Salam and Glashow which provided a unified description of the electromagnetic and weak interactions within the frame work of Quantum Field Theory. In these considerations spacetime is still Minkowski spacetime, that is Special Relativity holds.\\
However there has been one problem whose solution has defied nearly a century of work. This is the unification of gravitation with electromagnetism, Einstein's unfulfilled dream. The solution now seems within sight-- in Quantum Super String theory and in Quantum Gravitation approaches, including the author's own formulation. However there is a price to be paid-- we have to make a break with the past and introduce another counterintuitive concept, namely that spacetime is not a continuum, or is not a differentiable manifold, but rather is in some sense, itself quantized. This is a major departure from the Minkowski Spacetime of Special Relativity and Quantum Field Theory and also the Reimannian Spacetime of General Relativity. We will touch upon this in the sequel.
\section{Renormalization and Fuzzy Spacetime}
Let us start with the problem of renormalization, which as is well known was encountered first in Classical Electrodynamics \cite{rohr}. This was because the electromagnetic self energy of an electron viz., $\frac{e^2}{r}$ would $\to \infty$ as the size $r$ of the electron $\to 0$. On the other hand if $r$ were not to $\to 0$, that is the electron had a finite size, then this would lead to its own problems requiring the introduction of, for example Poincare stresses to hold the electron together (Cf.\cite{rohr}).\\
Let us consider the electron as having a bare mass and a physical mass, that is \cite{hooft}
\begin{equation}
m_{\mbox{phys}} = m_{\mbox{bare}} + \frac{e^2}{r}\label{ea}
\end{equation}
Then we could still have in the limit $r \to 0$, a finite physical mass, which would be what is actually measured, by allowing the infinite two terms on the right side of (\ref{ea}) to cancel each other out. This means we could preserve Special Relativity and at the same time recover a finite physical mass.\\
However on closer analysis, in Quantum Theory, there is no real problem with Special Relativity and superluminal or non-local velocities if $r$ were to be non zero, but of the order of the Compton wavelength $l$. We can see this as follows. A particle can travel from the spacetime point $x_1$ to the spacetime point $x_2$ causally only if the interval is time like, that is
$$\eta_{\alpha \beta} (x_1 - x_2)^\alpha (x_1 - x_2)^\beta < 0$$
On the other hand because of the Uncertainty Principle there is a non zero probability for a particle to move from $x_1$ to $x_2$ even if the interval is space like, that is with superluminal velocity as long as
$$(x_1 - x_2)^2 - (x_1^0 - x_2^0)^2 \leq \frac{h^2}{m^2} (c = 1)$$
In other words there is a breakdown of causal physics within the Compton scale 
 \cite{wein}.\\
All this illustrates the limitations of point spacetime and has direct relevance to our discussion in subsequent sections where we will introduce the idea of a minimum physical scale $l,\tau$ and fuzzyness. 
\section{Action at a Distance Electrodynamics}
We begin with classical electrodynamics. From a classical point of view a charge that is accelerating radiates energy which dampens its motion. This is given by the well known Maxwell-Lorentz equation, which in units $c = 1$, \cite{hoyle}, and $\tau$ being the proper time, while $\imath = 1,2,3,4$, is,
\begin{equation}
m \frac{d^2x^\imath}{d\tau^2} = e F^{\imath k} \frac{dx^k}{d\tau} + \frac{4e}{3} g_{\imath k} \left(\frac{d^3x^\imath}{d\tau^3} \frac{dx^l}{d\tau} - \frac{d^3x^l}{d\tau^3} \frac{dx^\imath}{d\tau}\right) \frac{dx^k}{d\tau},\label{e1}
\end{equation}
The first term on the right is the usual external field while the second term is the damping field which is added ad hoc by the requirement of the energy loss due to radiation. In 1938 Dirac introduced instead of (\ref{e1}),
\begin{equation}
m \frac{d^2x^\imath}{d\tau^2} = e \left\{F^\imath_k + R^\imath_k\right\} \frac{dx^k}{d\tau}\label{e2}
\end{equation}
where
\begin{equation}
R^\imath_k \equiv \frac{1}{2} \left\{F^{\mbox{ret}\imath}_k - F^{\mbox{adv}\imath}_k\right\}\label{e3}
\end{equation}
In (\ref{e3}), $F^{\mbox{ret}}$ denotes the retarded field and $F^{\mbox{adv}}$ the advanced field. While the former is the causal field where the influence of a charge at $A$ is felt by a charge at $B$ at a distance $r$ after a time $t = \frac{r}{c}$, the latter is the advanced acausal field which acts on $A$ from a future time. 
In effect what Dirac showed was that the radiation damping term in (\ref{e1}) or (\ref{e2}) is given by (\ref{e3}) in which an antisymmetric difference of the advanced and retarded fields is taken, which of course seemingly goes against causality as the advanced field acts from the future backwards in time. It must be mentioned that Dirac's prescription lead to the so called runaway solutions, with the electron acquiring larger and larger velocities in the absense of an external force. This he related to the infinite self energy of the point electron as can be seen from the second term on the right side of (\ref{ea}).\\
As far as the breakdown of causality is concerned, this takes place within a period $\sim \tau$, the Compton time. It was at this stage that Wheeler and Feynman reformulated the above action at a distance formalism in terms of what has been called their Absorber Theory. In their formulation, the field that a charge would experience because of its action at a distance on the other charges of the universe, which in turn would act back on the original charge is given by
\begin{equation}
Re = \frac{2e^2}{3} \frac{d}{dt} (\ddot{\bf x})\label{e4}
\end{equation}
The interesting point is that instead of considering the above force in (\ref{e4}) at the charge $e$, if we consider the responses in its neighbourhood, in fact a neighbourhood at the Compton scale, as was argued recently by the author \cite{iaad}, the field would be precisely the Dirac field given in (\ref{e2}) and (\ref{e3}). The net force emanating from the charge is thus given by
\begin{equation}
F^{\mbox{ret}} = \frac{1}{2} \left\{ F^{\mbox{ret}} + F^{\mbox{adv}}\right\} + \frac{1}{2} \left\{F^{\mbox{ret}} - F^{\mbox{adv}}\right\}\label{e5}
\end{equation}
which is the acceptable causal retarded field. The causal field now consists of the time symmetric field of the charge $e$ together with the Dirac field, that is the second term in (\ref{e5}), which represents the response of the rest of the charges. Interestingly in this formulation we have used a time symmetric field, viz., the first term of (\ref{e5}) to recover the retarded field with the correct arrow of time.\\
There are two important inputs which we can see in the above formulation. The first is the action of the rest of the universe at a given charge and the other is spacetime intervals which are of the order of the Compton scale. Infact we can push the above calculations further. The work done on a charge $e$ at $O$ by the charge at $P$ a distance $r$ away in causing a displacement $dx$ is given by (ignoring a cosine factor which merely gives a small numerical factor), 
\begin{equation}
\frac{e^2}{r^2} dx\label{e6}
\end{equation}
Now the number of particles at distance $r$ from $O$ is given by
\begin{equation}
n(r) = \rho(r) \cdot 4\pi r^2\label{e7}
\end{equation}
where $\rho(r)$ is the density of particles. So using (\ref{e7}) in (\ref{e6}) the total work is given by
\begin{equation}
E = \int \int \frac{e^2}{r^2} 4\pi r^2 \rho dx dr\label{e8}
\end{equation}
which can be shown using a uniform average density $\rho$, to be $\sim mc^2$. We thus recover in (\ref{e8}) the inertial energy of the particle in terms of its electromagnetic interactions with the rest of the universe in an action at a distance scheme.\\
Interestingly this can also be deduced in the context of gravitation: The work done on a particle of mass $m$ which we take to be a pion, a typical elementary particle, by the rest of the particles (pions) in the universe is given by
\begin{equation}
\frac{Gm^2N}{R}\label{e9}
\end{equation}
It is known that in (\ref{e9}) $N \sim 10^{80}$ while $R \sim \sqrt{N}l$, the well known Weyl-Eddington formula. Whence the gravitational energy of the pion is given by
\begin{equation}
\frac{Gm^2\sqrt{N}}{l} = \frac{e^2}{l} \sim mc^2\label{e10}
\end{equation}
where in (\ref{e10}) we have used the fact that
\begin{equation}
Gm^2 \sim \frac{e^2}{\sqrt{N}}\label{e11}
\end{equation}
(It must be mentioned that though the Eddington formula and (\ref{e11}) were empirical, they can infact be deduced from theory \cite{cu}, as we will see shortly.) 
\section{The Machian Universe}
This dependence of the mass of a particle on the rest of the universe was argued by Mach in the nineteenth century itself in what is now famous as Mach's Principle \cite{mwt,jv}. The Principle is counterintuitive in that we tend to consider the mass which represents the quantity of matter in a particle to be an intrisic property of the particle. But the following statement of Mach's Principle shows it to be otherwise thus going counter to ideas of locality and causality.\\
If there were no other particles in the universe, then the force acting on the particle $P$ would vanish and so we would have by Newton's second law
\begin{equation}
m\vec{a} = O\label{e12}
\end{equation}
Can we conclude that the acceleration of the particle vanishes? Not if we do not postulate the existence of an absolute background frame in space. In the absense of such a Newtonian absolute space frame, the acceleration $\vec{a}$ would infact be arbitrary, because we could measure this acceleration with respect to arbitrary frames. Then (\ref{e12}) implies that $m = 0$. That is, in the absense of any other matter in the universe, the mass of a material particle would vanish. From this point of view the mass of a particle depends on the rest of the material content of the universe. This has been brought out by the above calculations in (\ref{e8}) and (\ref{e10}).\\
Though Einstein was an admirer of Mach's ideas, his Special Theory of Relativity went counter to them. He subscribed as noted, to the concept of locality according to which information about a part of the universe can be obtained by dealing with that part alone and without taking into consideration the rest of the universe at the same time. In his words, \cite{singh} ``But one supposition we should, in my opinion absolutely hold fast: the real factual situation of the system $S_2$ is independent of what is done with the system $S_1$ which is spatially separated from the former.''Further, causality is another cornerstone in Einstein's Physics.
\section{The Quantum Universe}
The advent of Quantum Mechanics however threw up as noted several counter intuitive ideas and Einstein could not reconcile to them. One of these ideas was the wave particle duality. Another was that of the collapse of the wave function in which process causality becomes a casuality. To put it simply, if the wave function is a super position of the eigen states of an observable, then a measurement of the observable yields one of the eigen values no doubt, but it is not possible to predict which one. Due to the act of observation, the wave function instantly collapses to any one of its eigen states in an acausal manner. To put it another way, the wave function obeys the causal Schrodinger equation, for example, till the instant of observation at which point, causality ceases.\\
Another important counter intuitive feature of Quantum Mechanics is that of non locality. In fact Einstein with Podolsky and Rosen put forward in 1935 his arguments for the incompleteness of Quantum Mechanics on this score \cite{singh,EPR}. This has later come to be known as the EPR paradox. To put it in a simple way, without sacrificing the essential concepts, let us consider two elementary particles, for example two protons kept together somehow. They are then released and move in opposite directions. When the first proton reaches the point $A$ its momentum is measured and turns out to be say, $\vec{p}$. At that instant we can immediately conclude, without any further measurement that the momentum of the second proton which is at the point $B$ is $-\vec{p}$. This follows from the Conservation of Linear Momentum, and is perfectly acceptable in Classical Physics, in which the particles possess a definite momentum at each instant.\\
In Quantum Physics, the difficulty is that we cannot know the momentum at $B$ until and after a measurement is actually performed, and then that value of the momentum is unpredictable. What the above experiment demonstrates is that the proton at $B$ instantly came to have the value $-\vec{p}$ for its momentum without any further measurement, when the momentum of the proton at $A$ was measured. This ``instant'' or ``spooky action at a distance'' feature was unacceptable to Einstein.\\
In Quantum Theory however this is legitimate because of another counter intuitive feature which is called Quantum Nonseparability. That is, if two systems interact and then separate to a distance, they still have a common state vector. This goes against the concept of locality and causality, because it implies instantaneous interaction between distant systems. So in the above example, even though the protons at $A$ and $B$ may be separated, they still have a common wave function which collapses with the measurement of the momentum of any one of them and selfconsistently provides an explanation. This nonseparability has been characterised by Schrodinger in the following way: ``I would not call that \underline{one}, but rather \underline{the} characteristic of Quantum Mechanics.'' For Einstein however this was like spooky action at a distance. All this has been experimentally verified since 1980 which sets at rest Einstein's objections.\\
However this ``entanglement'' as it is called these days, between distant objects in the universe, does not really manifest itself though it is perfectly legitimate and observable in a universe that consists of let us say just two particles. But a measurement destroys the entanglement. Now in the universe at large as there are so many particles and correspondingly a huge amount of interference, the entanglement is considerably weakened. This was the crux of Schrodinger's arguments. What is these days called decoherence works along these lines. This is in fact the explanation for the famous ``Schrodinger's Cat'' paradox.\\
This paradox can be explained in the following simple terms: A cat is in an enclosure along with, let us say a microscopic amount of radioactive material. If this material decays, emitting let us say an electron, the electron would fall on a vial of cyanide, releasing it and killing the cat in the process. Let us say that there is a certain probability of such an electron being emitted. So there is the same probability for the cat to be killed. There is also a probability that the electron is not emitted, so that there is the same probability for the cat to remain alive. The cat is therefore in a state which is a superposition of the alive and dead states. It is only when an observer makes an observation that this superposed wave function collapses into either the dead cat state or the alive and kicking cat state, and this happening is acausal. So it is only on an observation being made that the cat is killed or saved, and that too in an unpredictable manner. Till the observation is made the cate is described by the superposed wave function and is thus neither alive nor dead.\\
The resolution of this paradox-- it is a paradox-- is of course quite simple. The paradox is valid if the system consists of such few particles and at such distances that they do not interact with each other. Clearly in the real world this idealization is not possible. There are far too many particles and interferences taking place all the time and the superposed wave function would have collapsed almost instantly. This role of the environment has come to be called decoherence. We will return to this point shortly.\\
The important point is that all of Classical and Quantum Physics is based on such idealized laws as if there were no interferences present, that is what may be called a two body scenario, is implicit. Clearly this is not a real life scenario.
\section{The Zero Point Field}
Another counter intuitive concept which Quantum Theory introduces is that of the Zero Point Field or Quantum Vacuum. If there were a vacuum, in which at a given point the momentum (and energy) would vanish, then by the Heisenberg Principle, the point itself becomes indeterminate. More realistically, in the vacuum the average energy vanishes but there are fluctuations-- these are the Zero Point Fluctuations. A more classical way of looking at this is that the source free vacuum electromagnetic equations have non zero solutions, in addition to the zero solutions. Interestingly we can argue that the Zero Point Field leads to a minimum interval at the Compton scale \cite{def}.\\
The momentum operators $m\vec{v}$ do not satisfy the Quantum Mechanical commutation relations with the position coordinators. But if we add the electromagnetic momentum due to the background Zero Point Field, then we recover the Quantum Mechanical commutation relations and spin half \cite{sachi}. Thus it appears that Classical Mechanics together with the Zero Point Field leads to Quantum Theory. Infact several authors like Marshall, Boyer and others had argued for Quantum Mechanics arising from stochastic electrodynamics \cite{depena2}.\\
Indeed we could even compute the energy due to the Zero Point Field, which in turn gives rise to the Lorentz force and recover the inertial energy.\\
The manifestation of the Zero Point Field has been experimentally tested in what is called the Lamb Shift, which is caused by the fact that the Zero Point Field buffets an ordinary electron in an atom. It has also been verified in the famous Casimir effect \cite{mes,mdef}. The Zero Point Field in this case manifests itself as an attractive force between two parallel plates.\\
Interestingly, based on such a Quantum Vacuum and the minimum spacetime intervals the author had proposed a cosmological model in 1997 which predicted an accelerating universe and  a small cosmological constant. In addition, several so called large number relations which had been written off as inexplicable empirical coincidences (including the Eddington formula and the electromagnetism gravitation strengths ratio, alluded to) were shown to follow from the theory \cite{ijmpa}. At that time the prevailing cosmological model was one of dark matter and a decelerating universe. Observational confirmation started coming for the new predictions from 1998 itself while the observational discovery of dark energy, which displaces dark matter, was the scientific Breakthrough of the Year 2003 of the American Association for Advancement of Science \cite{science}. Dark energy goes hand in hand with a cosmological constant of the kind Einstein proposed and then discarded.\\
It may be observed that the idea of the Zero Point Field was introduced as early as in 1911 by Max Planck himself to which he assigned an energy $\frac{1}{2} \hbar \omega$. Nernst, a few years later extended these considerations to fields and believed that there would be several interesting consequences in Thermodynamics and even Cosmology.\\
Infact later authors argued that there must be fluctuations of the Quantum Electromagnetic Flield, as required by the Heisenberg Principle, so that we have for an extent $\sim L$ ($B$ being the magnetic field),
\begin{equation}  
(\Delta B)^2 \geq \hbar c/L^4\label{e13}
\end{equation}
Whence from (\ref{e13}), the dispersion in energy in the entire volume $\sim L^3$ is given by
\begin{equation}
\Delta E \sim \hbar c/L\label{e14}
\end{equation}
(It should be noticed that if $L$ is the Compton wavelength, then (\ref{e14}) gives us the energy of the particle.)Interestingly Braffort and coworkers deduced the Zero Point Field from the Absorber Theory of Wheeler and Feynman, which we encountered earlier. In the process they found that the spectral density of the vacuum field was given by \cite{depena}
\begin{equation}
\rho (\omega) = \mbox{const}\cdot \omega^3\label{e15}
\end{equation}
There have been other points of view which converge to the above conclusions. In any case as we have seen a little earlier, it would be too much of an idealization to consider an atom or a charged particle to be an isolated system. It is interacting with the rest of the universe and this produces a random field.\\
It has also been shown that the constant of proportionality in (\ref{e15}) is given by (Cf.ref.\cite{depena})
$$\frac{\hbar}{2\pi^2 c^3}$$
Interestingly such a constant is implied by Lorentz invariance.\\
From the point of view of Quantum Electrodynamics we reach conclusions similar to those seen above. As Feynman and Hibbs put it \cite{fh} ``Since most of the space is a vacuum, any effect of the vacuum-state energy of the electromagnetic field would be large. We can estimate its magnitude. First, it should be pointed out that some other infinities occuring in quantum-electrodynamic problems are avoided by a particular assumption called the \underline{cutoff rule}. This rule states that those modes having very high frequencies (short wavelength) are to be excluded from consideration. The rule is justified on the ground that we have no evidence that the laws of electrodynamics are obeyed for wavelengths shorter than any which have yet been observed. In fact, there is a good reason to believe that the laws cannot be extended to the short-wavelength region.\\
``Mathematical representations which work quite well at longer wavelengths lead to divergences if extended into the short wavelength region. The wavelengths in question are of the order of the Compton wavelength of the proton; $1/2\pi$ times this wavelength is $\hbar/mc \simeq 2 \times 10^{-14}cm$.\\
``For our present estimate suppose we carry out sums over wave numbers only up to the limiting value $k_{max} = mc/\hbar$. Approximating the sum over states by an integral, we have, for the vacuum-state energy per unit volume,
$$\frac{E_e}{\mbox{unit \, vol}} = 2 \frac{\hbar c}{2(2\pi)^3} \int^{k_{max}}_0 k 4\pi k^2 dk = \frac{\hbar c k^4_{max}}{8\pi^2}$$
``(Note the first factor $2$, for there are two modes for each $k$). The equivalent mass of this energy is obtained by dividing the result by $c^2$. This gives
$$\frac{m_0}{\mbox{unit \, vol}} = 2 \times 10^{15} g/cm^3$$
Such a mass density would, at first sight at least, be expected to produce very large gravitational effects which are not observed. It is possible that we are calculating in a naive manner, and, if all of the consequences of the general theory of relativity (such as the gravitational effects produced by the large stresses implied here) were included, the effects might cancel out; but nobody has worked all this out. It is possible that some cutoff procedure that not only yields a finite energy density for the vacuum state but also provides relativistic invariance may be found. The implications of such a result are at present completely unknown.\\
``For the present we are safe in assigning the value zero for the vacuum-state energy density. Up to the present time no experiments that would contradict this assumption have been performed.''\\
However the high density encountered above is perfectly meaningful if we consider the Compton scale cut off $\sim 10^{-13}cm$: Within this volume the density gives us back the mass of an elementary particle like the pion. All this can be put into perspective in the following way.\\
It has been shown in detail by the author that the universe can be considered to have an underpinning of ZPF oscillators at the Planck scale \cite{bgsfpl}. Indeed in all recent approaches towards a unified formulation of gravitation and electromagnetism (including String Theory), the differentiable spacetime manifold of Classical Physics and Quantum Physics has been abandoned as noted earlier and we consider the minimum Planck scale $l_P \sim 10^{-33}cms$ and $\tau_P \sim 10^{-42}secs$ \cite{uof}. We can then show that the universe is a coherent mode of $\bar{N} \sim 10^{120}$ Planck oscillators, spaced a distance $l_P$ apart, that is at the Planck scale. Then the spatial extent is given by
\begin{equation}
R = \sqrt{\bar{N}}l_P\label{e16}
\end{equation}
The mass of the universe is given by
\begin{equation}
M = \sqrt{\bar{N}} m_P\label{e17}
\end{equation}
where $m_P$ is the Planck mass. Moreover we can show that a typical elementary particle like the pion is the ground state of $n \sim 10^{40}$ oscillators and we have (Cf.ref.\cite{bgsfpl})
\begin{equation}
m = \frac{m_P}{\sqrt{n}}\label{e18}
\end{equation}
\begin{equation}
l = \sqrt{n} l_P\label{e19}
\end{equation}
There are $N \sim 10^{80}$ such elementary particles in the universe. Whence we have
\begin{equation}
M = Nm\label{e20}
\end{equation}
We note that equations like (\ref{e16}) and (\ref{e19}) have the Brownian Random Walk character. At this stage we see asymmetry between equations (\ref{e17}), (\ref{e18}) and (\ref{e20}). The reason is that the universe is an excited state of $\bar{N}$ oscillators whereas an elementary particle is a stable ground state of $n$ Planck oscillators. Furthermore, let us denote the state of each Planck oscillator by $\phi_n$; then the state of the universe can be described in the spirit of entanglement discussed earlier by
\begin{equation}
\psi = \sum_{n} c_n \phi_n,\label{e21}
\end{equation}
$\phi_n$ can be considered to be eigen states of energy with eigen values $E_n$. It is known that (\ref{e21}) can be written as \cite{bgscsf}
\begin{equation}
\psi = \sum_{n} b_n \bar{\phi}_n\label{e22}
\end{equation}
where $|b_n|^2 = 1 \, \mbox{if}\, E < E_n < E + \Delta$ and $= 0$ otherwise under the assumption
\begin{equation}
\overline{(c_n,c_m)} = 0, n \ne m\label{e23}
\end{equation}
(Infact $n$ in (\ref{e23}) could stand for not a single state but for a set of states $n_\imath$, and so also $m$). Here the bar denotes a time average over a suitable interval. This is the well known Random Phase Axiom and arises due to the total randomness amongst the phases $c_n$. Also the expectation value of any operator $O$ is given by
\begin{equation}
< O > = \sum_{n} |b_n|^2 (\bar{\phi}_n, O \bar{\phi}_n)/\sum_{n} |b_n|^2\label{e24}
\end{equation}
Equations (\ref{e22}) and (\ref{e24}) show that effectively we have incoherent states $\bar{\phi}_1, \bar{\phi}_2,\cdots$ once averages over time intervals for the phases $c_n$ in (\ref{e23}) vanish owing to their relative randomness.\\
In the light of the preceding discussion of random fluctuations, we can interpret all this meaningfully: We can identify $\phi_n$ with the ZPF. The time averages are the same as Dirac's zitterbewegung averages over intervals $\sim \frac{\hbar}{mc^2}$ (Cf.ref.\cite{cu}). We then get disconnected or incoherent particles from a single background of vacuum fluctuations exactly as before. The incoherence arises because of the well known random phase relation (\ref{e23}), that is after averaging over the suitable interval. Here the entanglement is weakened by the interactions and hence we have (\ref{e20}) for elementary particles, rather than (\ref{e17}).\\
How do we characterize time in this scheme? To consider this problem, we observe that the ground state of $\bar{N}$ Planck oscillators considered above would be, exactly as in (\ref{e18}),
\begin{equation}
\bar {m} = \frac{m_P}{\sqrt{\bar{N}}} \sim 10^{-65}gms\label{ex2}
\end{equation}
It is interesting that it follows from thermodynamic arguments that this is the smallest possible observable mass in the universe. 
The universe is an excited state and consists of $\bar{N}$ such ground state
levels and so we have, from (\ref{ex2})
$$M = \bar{m} \bar{N} = \sqrt{\bar{N}} m_P \sim 10^{55}gms,$$ 
as required, $M$ being the mass of the universe. As can be easily calculated, the Compton wavelength and time of $\bar{m}$ turn out to be the radius and age of the universe.\\ 
Due to the fluctuation $\sim
\sqrt{n}$ in the levels of the $n$ oscillators making up an elementary
particle, the energy is, remembering that $mc^2$ is the ground state,
$$\Delta E \sim \sqrt{n} mc^2 = m_P c^2,$$ 
 and so the indeterminacy time is
$$\frac{\hbar}{\Delta E} = \frac{\hbar}{m_Pc^2} = \tau_P,$$ as indeed
we would expect.\\ 
The corresponding minimum indeterminacy length
would therefore be $l_P$.  We thus recover the Planck scale. One of the consequences of the minimum
spacetime cut off-- whether it be in Quantum Super String theory or Quantum Gravity approaches or the author's own approach-- is that the Heisenberg Uncertainty
Principle takes an extra term. Thus we have,
\begin{equation}
\Delta x \approx \frac{\hbar}{\Delta p} + \alpha \frac{\Delta
p}{\hbar},\, \alpha = l^2 (\mbox{or}\, l^2_P)\label{ex6}
\end{equation}
where $l$ (or $l_P$) is the minimum interval under consideration (Cf.\cite{cu,uof}). 
The first term gives the usual Heisenberg Uncertainty
Principle.\\
However, the second and extra term in (\ref{ex6}) has the implication that as we go down to arbitrarily small distances, we end up once again at the large scale. (This is sometimes called duality in String theory.) This not only prohibits arbitrarily small resolution, but shows up the universe as some sort of a four dimensional Klein's bottle.\\
Application of the time analogue of (\ref{ex6}) for the
indeterminacy time $\Delta t$ for the fluctuation in energy $\Delta
\bar{E} = \sqrt{N} mc^2$ in the $N$ particle states of the universe
gives exactly as above,
$$\Delta t = \frac{\Delta E}{\hbar} \tau^2_P =
\frac{\sqrt{N}mc^2}{\hbar} \tau^2_P = \frac{\sqrt{N}
m_Pc^2}{\sqrt{n}\hbar} \tau^2_P = \sqrt{n} \tau_P = \tau$$ 
In other words, for the fluctuation
$\sqrt{N}$, the time is $\tau$. It must be re-emphasized that the
Compton time $\tau$ of an elementary particle, is an interval within
which there are unphysical effects like zitterbewegung - as pointed
out by Dirac, it is only on averaging over this interval, that we
return to meaningful Physics.  To continue, we then have,
\begin{equation}
dN/dt = \sqrt{N}/\tau\label{ex3}
\end{equation}
On the other hand
due to the fluctuation in the $\bar{N}$ oscillators
constituting the universe, the fluctuational energy is similarly given
by
$$\sqrt{\bar {N}} \bar {m} c^2,$$ which is the same as (\ref{ex2})
above. Another way of deriving (\ref{ex3}) is to observe that as
$\sqrt{n}$ particles appear fluctuationally in time $\tau_P$ which is,
in the elementary particle time scales, $\sqrt{n} \sqrt{n} = \sqrt{N}$
particles in $\sqrt{n} \tau_P = \tau$. That is, the rate of the
fluctuational appearance of particles is
$$
\left(\frac{\sqrt{n}}{\tau_P}\right) = \frac{\sqrt{N}}{\tau} = dN/dt$$
which is (\ref{ex3}). From here by integration,
$$T = \sqrt{N} \tau$$ $T$ is the time elapsed from $N = 1$ and $\tau$
is the Compton time. This gives $T$ its origin in the fluctuations -
there is no smooth ``background'' (or ``being'') time - the root of
time is in ``becoming''. It is the time of a Brownian Wiener
process: A step $l$ gives a step in time $l/c \equiv \tau$ and
therefore the Brownian relation $\Delta x = \sqrt{N} l$ gives $T = \sqrt{N} \tau$ (Cf.refs.\cite{bgsfpl} and \cite{uof}). Time is
born out of acausal fluctuations which are random and therefore
irreversible. Indeed, there is no background time. States are created and states are destroyed, but the net result is the creation of $\sqrt{N}$ states and time is proportional to $\sqrt{N}$, $N$ being the number of particles which
are being created spontaneously from the ZPF by fluctuations to the higher energy states of the coherent $\bar{N}$ Planck oscillators.
\section{The Underpinning of the Universe}
So our description of the universe at the Planck scale is that of an entangled wave function as in (\ref{e21}). However we percieve the universe at the elementary particle or Compton scale, where the random phases would have weakened the entanglement, and we have the description as in (\ref{e22}) or (\ref{e24}). Does this mean that the $N$ elementary particles in the universe are totally incoherent in which case we do not have any justification for treating them to be in the same spacetime? We can argue that they still interact amongst each other though in comparison this is ``weak''. For instance let us consider the background ZPF whose spectral frequency is given by (\ref{e15}). If there are two particles at $A$ and $B$ separated by a distance $r$, then those wavelengths of the ZPF which are atleast $\sim r$ would connect or link the two particles. Whence the force of interaction between the two particles is given by, remembering that $\omega \propto \frac{1}{r}$,
\begin{equation}
\mbox{Force}\, \quad  \propto \int^\infty_r \omega^3 dr \propto \frac{1}{r^2}\label{e27}
\end{equation}
Thus from (\ref{e27}) we are able to recover the familiar Coulomb Law of interaction. The background ZPF thus enables us to recover the action at a distance formulation. Infact a similar argument can be given \cite{fisch} to recover from QED the Coulomb Law--here the carriers of the force are the virtual photons, that is photons whose life time is within the Compton time of uncertainty permitted by the Heisenberg Uncertainty Principle.\\
It is thus possible to synthesize the field and action at a distance concepts, once it is recognized that there are minimum spacetime intervals at the Compton scale \cite{iaad}. Many of the supposed contradictions arise because of our characterization in terms of spacetime points and consequently a differentiable manifold. Once the minimum cut off at the Planck scale is introduced, this leads to the physical Compton scale and a unified formulation free of divergence problems.\\
It must be stressed that this view of spacetime is not only that of a non differentiable manifold but also has another characteristic: spacetime rather than being a container or stage, is a result of its constituents, a view propounded by Liebniz. It is as if the actors make up the stage itself, another apparently counterintuitive concept. We now make a few comments.\\
We had seen that the Dirac formulation of Classical Electrodynamics needed to introduce the acausal advanced field in (\ref{e3}). However the acausality was again within the Compton time scale. Infact this fuzzy spacetime can be modelled by a Wiener process as discussed in \cite{uof}(Cf. also \cite{nottale}). The point here is that the backward and forward time derivatives for $\Delta t \to 0^-$ and $0^+$ respectively do not cancel, as they should not, if time is fuzzy. So we automatically recover from the electromagnetic potential the retarded field for forward derivatives and the advanced fields for backward derivatives. In this case we have to consider both these fields. Causality however is recovered as in (\ref{e5}). This is a transition to intervals which are greater in magnitude compared to the Compton scale in which case the latter can be neglected.\\
It must also be mentioned that a few assumptions are implicit in the conventional theory using differentiable spacetime manifolds. In the variational problem we use the conventional $\delta$ (variation) which commutes with the time derivatives. So such an operator is constant in time. So also the energy momentum operators in Dirac's displacement operators theory are the usual time and space derivatives of Quantum Theory. But here the displacements are ``instantaneous''. They are valid in a stationary or constant energy scenario, and it is only then that the space and time operators are on the same footing as required by Special Relativity \cite{davydov}. Infact it can be argued that in this theory we neglect intervals $\sim 0(\delta x^2)$ but if $\delta x$ is of the order of the Compton scale and we do not neglect the square of this scale, then the space and momentum coordinates become complex indicative of a noncommutative geometry which has been discussed in detail \cite{bgscst,bgsknmetric,uof}. What all this means is that it is only on neglecting $0(l^2)$ that we have the conventional spacetime of Quantum Theory, including relativistic Quantum Mechanics and Special Relativity, that is the Minkowski spacetime.\\
Coming to the conservation laws of energy and momentum these are based on translation symmetries \cite{roman}-- what it means is the operators $\frac{d}{dx} \, \mbox{or}\,  \frac{d}{dt}$ are independent of $x$ and $t$. There is here a homogeneity property of spacetime which makes these laws non local. This has to be borne in mind, particularly when we try to explain the EPR paradox.\\
The question how a ``coherent'' spacetime can be extracted out of the particles of the universe could be given a mathematical description along the following lines: Let us say that two particles $A$ and $B$ are in a neighbourhood, if they interact at any time. We also define a neighbourhood of a point or particle $A$ as a subset of all points or particles which contains $A$ and at least one other point. If a particle $C$ interacts with $B$ that is, is in a neighbourhood of $B$, then we would say that it is also in the neighbourhood of $A$. That is we define the transitivity property for neighbourhoods. We can then assume the following property \cite{bgsaltaisky}:\\
Given two distinct elements (or even subsets) $A$ and $B$, there is a neighbourhood $N_{A_1}$ such that $A$ belongs to $N_{A_1}$, $B$ does not belong to $N_{A_1}$ and also given any $N_{A_1}$, there exists a neithbourhood $N_{A_\frac{1}{2}}$ such that $A \subset N_{A_\frac{1}{2}} \subset N_{A_1}$, that is there exists an infinite sequence of neithbourhoods between $A$ and $B$. In other words we introduce topological ``closeness''. Alternatively, we could introduce the reasonable supposition that these are a set of Borel subsets.\\
From here, as in the derivation of Urysohn's lemma \cite{simmons}, we could define a mapping $f$ such that $f(A) = 0$ and $f(B) = 1$ and which takes on all intermediate values. We could now define a metric, $d(A,B) = |f(A) - f(B)|$. We could easily verify that this satisfies the properties of a metric.\\
It must be remarked that the metric turns out to be again, a result of a global or a series of large sets, unlike the usual local picture in which it is the other way round.    
\section{The Path Integral Formulation}
We first argue that the alternative Feynman Path Integral formulation essentially throws up fuzzy spacetime. To recapitulate \cite{fh,nottale,it}, if a path is given by
$$x = x(t)$$
then the probability amplitude is given by
$$\phi (x) = e^{\imath \int^{t_2}_{t_1}L(x, \dot{x})dt}$$
So the total probability amplitude is given by
$$\sum_{x(t)} \phi (x) = \sum e^{\imath \int^{t_2}_{t_1}L(x,\dot{x})dt} \equiv \sum e^{\frac{\imath}{\hbar}S}$$
In the Feynman analysis, the path
$$x = \bar{x} (t)$$
appears as the actual path for which the action is stationery. For paths very close to this, there is constructive interference, whereas for paths away from this the interference is destructive.\\
We notice that this is also the formulation of the random phase encountered earlier \cite{huang}. However it is well known that the convergence of the integrals requires the Lipshitz condition viz.,
\begin{equation}
\Delta x^2 \approx a \Delta t\label{e1b}
\end{equation}
We could say that only those paths satisfying (\ref{e1b}) constructively interfere. We would now like to observe that (\ref{e1b}) is the well known Brownian or Diffusion equation related to our earlier discussion of the Weiner process. This has been commented upon extensively \cite{heap,cu,uof} and also \cite{rief}. The point is that (\ref{e1b}) again implies a minimum spacetime cut off, as indeed was noted by Feynman himself \cite{fh}, for if $\Delta t$ could $\to 0$, then the velocity would $\to \infty$.\\
As Feynman and Hibbs put it, ``... these irregularities are such that the ``average'' square velocity does not exist, where we have used the classical analogue in referring to an ``average.''\\
``If some average velocity is defined for a short time interval $\Delta t$, as, for example, $[x(t + \Delta t) - x(t)]/\Delta t$, the ``mean'' square value of this is $-\hbar /(\imath m \Delta t)$. That is, the ``mean'' square value of a velocity averaged over a short time interval is finite, but its value becomes larger as the interval becomes shorter.\\
It appears that quantum-mechanical paths are very irregular. However, these irregularities average out over a reasonable length of time to produce a reasonable drift, or ``average'' velocity, although for short intervals of time the ``average'' value of the velocity is very high...''\\
To put it another way we are taking averages over an interval $\Delta t$, within which there are unphysical processes. It is only after the average is taken, that we recover physical spacetime \cite{bgscsf}. If in the above, $\Delta t$ is taken as the Compton time, then we recover for the root mean squared velocity, the velocity of light.\\
As has been argued in detail \cite{cu,uof} this is exactly the situation which we encounter in the Dirac theory of the electron. There we have the unphysical zitterbewegung effects within the Compton time $\Delta t$ and as $\Delta t \to 0$ the velocity of the electron tends to the maximum possible velocity, that of light \cite{dirac}. It is only after averaging over the Compton scale that we recover meaningful physics.\\
This  existence of a minimum spacetime scale, it has been argued for quite sometime is the origin of fuzzy spacetime, described by a noncommutative geometry, consistent with Lorentz invariance. This was shown a long time ago by Snyder \cite{snyder}. In this case we have commutative relations like
$$[x_\imath , x_j] = \Theta_{\imath j} O(l^2)$$
\begin{equation}
[x_\imath , p_j] = \tilde{\Theta}_{\imath j} \hbar [1 + O(l^2)]\label{e2b}
\end{equation}
It is interesting to note that the momentum position commutation relations lead to the usual Quantum Mechanical commutation relations in the usual (commutative) spacetime if $O(l^2)$ is neglected where $l$ defines the minimum scale. Indeed, as noted elsewhere, we have at the smallest scale, a quantum of area. Another way of looking at this is to observe that the Quantum Mechanical path has the fractal dimension 2 (Cf.ref.\cite{uof}), corresponding to the Quantum of area. It is this ``fine structure'' of spacetime which is expressed by the noncommutative structure (\ref{e2b}). Neglecting $O(l^2)$ is equivalent to neglecting the above and returning to usual spacetime. (In other words Snyder's purely classical considerations at a Compton scale lead to Quantum Mechanics.
\section{A Degenerate Bose Assembly Model}
We now observe that the coherent $N'$ Planck oscillators referred to above could be considered to be a degenerate Bose assembly. In this case as is well known we have $(z \approx 1)$
$$v = \frac{V}{N}$$
(Cf.ref.\cite{huang}). $V$ the volume of the universe $\sim 10^{84}cm^3$. Whence
$$v = \frac{V}{N'} \sim 10^{-36}$$
So that the wavelength 
\begin{equation}
\lambda \sim (v)^{1/3} \sim 10^{-12}cm  = l\label{e3b}
\end{equation}
What is very interesting is that (\ref{e3b}) gives us the Compton length of a typical elementary particle like the pion. So from the Planck oscillators we are able to recover the elementary particles exactly as in the references \cite{psp,bgsfpl}.\\
Moreover, let us now consider the distant background assembly which is at nearly the same energy. In this case we have (Cf.ref.\cite{huang})
\begin{equation}
\langle n_{\vec{k}}\rangle = \frac{2}{e^{\beta h w}-1}\label{e4b}
\end{equation}
As we have assumed that the photons all have nearly the same energy, we have,
\begin{equation}
\langle n_{\vec {k}}\rangle = \langle n_{\vec{k}'}\rangle \delta (k - k')\label{e5b}
\end{equation}
where $\langle n_{\vec{k}'}\rangle$ is given by (\ref{e4b}), and $k \equiv |\vec{k}|$. The total number of photons $N$, in the volume $V$ being considered, can be obtained in the usual way,
\begin{equation}
N = \frac{V}{(2\pi)^3}\int^\infty_0 \, dk4\pi k^2\langle n_k\rangle\label{e6b}
\end{equation}
where $V$ is large. Inserting (\ref{e5b}) in (\ref{e6b}) we get,
\begin{equation}
N = \frac{2V}{(2\pi)^3} 4\pi k'^2 [\epsilon^\Theta - 1]^{-1} [k]\, \Theta \equiv \beta h w\label{e7b}
\end{equation}
where $[k]$ is a dimensionality constant of magnitude unity, introduced to compensate the loss of a factor $k$ in the integral (\ref{e6b}), owing to the $\delta$-function in (\ref{e5b}).\\
We observe that, $\Theta =hw/KT \approx 1$, since by (\ref{e5}), the photons have the same energy $hw$. We also use,
\begin{equation}
v = \frac{V}{N}, \lambda = \frac{2\pi c}{w} = \frac{2\pi}{k} \, \mbox{and}\, z = \frac{\lambda^3}{v}\label{e8b}
\end{equation}
$\lambda$ being the wavelength of the radiation. We now have from (\ref{e7}), using (\ref{e8b}),
$$(e - 1) = \frac{vk'^2}{\pi^2} [k] = \frac{8\pi}{k'}\frac{1}{z} [k]$$
At this stage we observe that as $z$ is dimensionless, this equation is perfectly consistent because both sides are dimensionless. Using (\ref{e8b}) we get:
\begin{equation}
z = \frac{8\pi}{k'(e-1)} = \frac{4\lambda}{(e-1)} [k]\label{e9b}
\end{equation}
From (\ref{e9b}) we conclude that, in this case,
\begin{equation}
\lambda = \frac{e-1}{4} = 0.4cm\label{e10b}
\end{equation} 
It must be observed that we consider the degenerate case in all the above considerations and so $z \approx 1$. What is very interesting is that (\ref{e10b}) this time gives us the correct microwave cosmic background radiation wavelength and temperature.\\
We can get an alternative justification for the above considerations. As shown in detail elsewhere \cite{anonwi}, if we balance the gravitational energy and Fermi energy of the cold cosmic background neutrinos and identify this with inertial energy of the neutrino,
\begin{equation}
\frac{GN_\nu m^2_\nu}{R} = \frac{N^{2/3}_\nu \hbar^2}{m_\nu R^2} = m_\nu c^2\label{e11b}
\end{equation}
we get the correct mass of the neutrino $m_\nu ,\sim 10^{-8}$ electron mass and the correct number of background neutrinos, $N_\nu \sim 10^{90}$.\\
We can then recover these neutrinos as a consequence of $N_\nu$ Planck oscillators considered above and also the Cosmic Microwave Background temperature of $\sim 3^\circ K$ (Cf. ref.\cite{anonwi} for details).\\
Yet another demonstration of all this is to start with the degenerate Fermi gas equation \cite{huang},
\begin{equation}
p^3_F = \hbar^3 (N/V)\label{e12b}
\end{equation}
If in (\ref{e12b}), $N = N_\nu$, above, we get back $m_\nu$ and the background temperature. Conversely, taking $T \sim 1^\circ K$, we can get $m_\nu$ and $N_\nu$. Thus a neutrino shows up as a cold electron, reminiscent of quarks and electrons being interchangeable at high temperatures.

\end{document}